\documentclass[english,A4paper,11pt]{article}
\usepackage[english]{babel}
\usepackage{graphicx,epsfig,epic} % standard LaTeX graphics tool
\usepackage{pifont}      % placeholder for figures

\usepackage{pict2e}
\usepackage{pifont}      % placeholder for figures
\usepackage{pict2e}
\usepackage{amssymb}
\usepackage{latexsym}
\usepackage{setspace}
\usepackage{amscd}
\def\pesp{\vskip .3cm}

\def\ni{\noindent}

\newtheorem{defn}{Definition}

\newtheorem{prop}{Proposition}
\newtheorem{prp}{Property}

\newtheorem{rmak}{Remark}

\begin{document}
\title{ Space-Time Entropy, Space of Singularities\\
 and Gravity Origin: a Case Study}
\author{\small Fay\c cal BEN ADDA{\footnote{Community College of Qatar, Doha, Qatar.
Email: f\_benaddafr@yahoo.fr}}}
\maketitle
\begin{abstract}
A new definition of entropy is introduced using a model that simulates an expanding space-time compatible with the fundamental principle of cosmology. The entropy is obtained by mean of a state function that measures the variation of the space-time normal curvature, from a highly compressed space to a lower compressed space. The defined entropy leads to work out a new understanding of the earliest conditions that last for a period estimated to 380 000 years after the Big Bang. It leads to understand via a short period of inflation the process that generates the uniform distribution of matter and energy at the surface of the last scattering. It involves gravitational singularities in a process of gradual decompression propitious to the incubation of matter recombination, and it allows to trace back gravity origin.\pesp



{\footnotesize {AMS}: 28D20 , 83C57 }

\end{abstract}

{\small\tableofcontents}

\section{Introduction}

The analysis of the Planck's CMB image has revealed that the distribution of mass/energy density of the universe is estimated today to 68,3 $\%$ of dark energy, 26,8 $\%$ of dark matter and 4,9$\%$ of normal matter (European Spacial Agency ESA /Planck). However, our understanding of the components of the distribution remains unknown to satisfactory define, in particular if many phenomena in relation with our universe are still unknown including the nature of the dark energy, the nature of the dark matter, the characteristics of the inflationary period, the universe phase transitions, the asymmetry in the average temperatures on opposite hemispheres of the sky (from CMB), and the earliest conditions of the universe expansion.\pesp

The oldest light propagation in the universe, known as the cosmic microwave background (CMB), is actually related to a non accessible young universe of 380 000 years old. A critical period is missing for the comprehension of the earliest conditions that generate the distribution of matter at large scale as well as its transformation together with the space-time expansion. During the 380 000 years, light was trapped within an opaque environment that left no traceable direct evidence to understand what happens in the earliest phase of the universe transformation. Many hypothetical attempts to explain the earliest conditions after the Big Bang have been elaborated and the list is not limited to the following selection (\cite{Caisar},\cite{Cai},\cite{Cook},\cite{Dai},\cite{Ellis},\cite{Haque},\cite{Haro},\cite{MarRin},\cite{Maity1},\cite{Maity2}).\pesp

To access the earliest conditions of the universe expansion, we introduce a new entropy that associates a state function sensitive to the variation of the space-time curvature, provided that this variation is directly related to its expansion. For this purpose we use a model introduced in \cite{BFP} that simulates an expanding space-time compatible with the fundamental principle of cosmology. The space-time is simulated by an accumulation of an infinite family of equal open balls endowed with a discrete simultaneous expansion in all directions.\pesp

\begin{figure}[!h]
\begin{minipage}[t]{7cm}
\centering
\includegraphics[width=7cm]{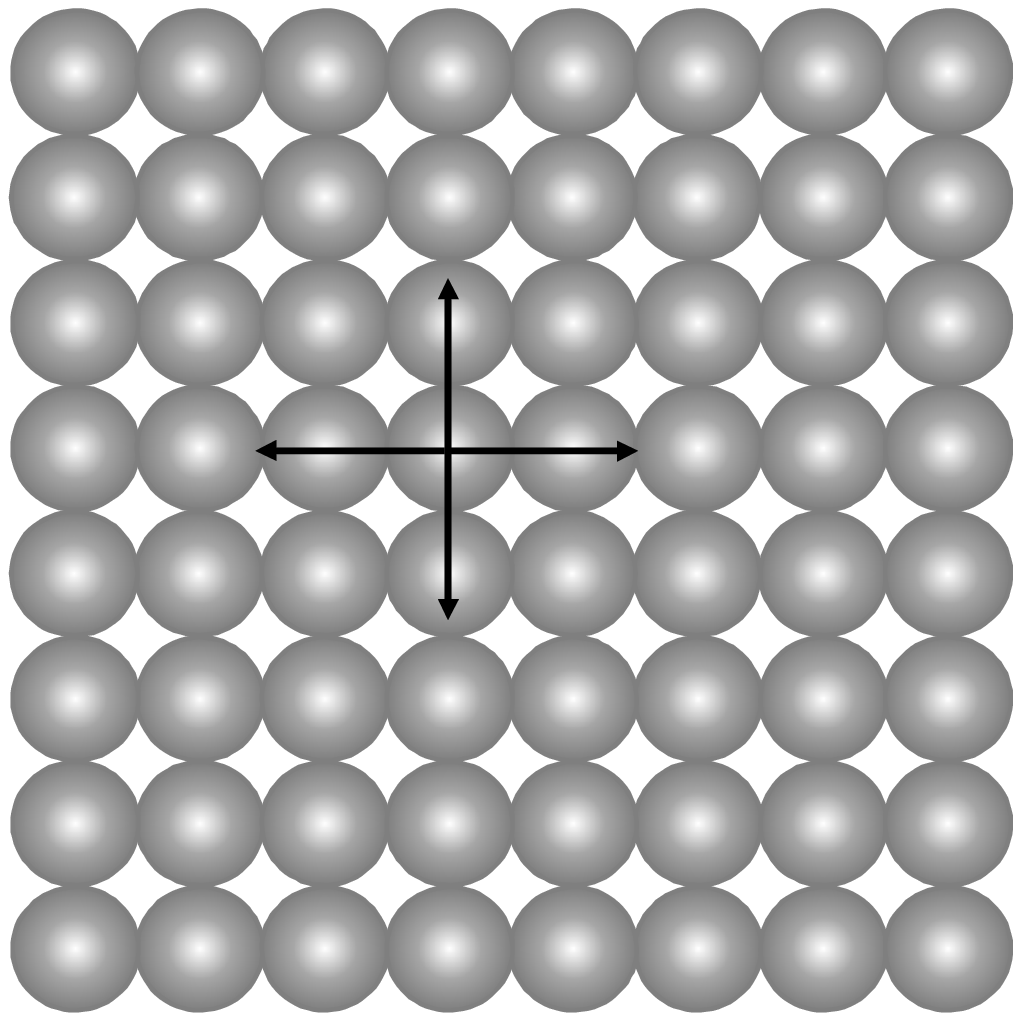}
\caption{\footnotesize Illustration of one layer of $8 \times 8$ equal ball packing, where each interstice is enclosed between four stacked balls: the uniformity is only verified within four orientations.}\label{Fig.1}
\end{minipage}
\hspace*{\fill}
\begin{minipage}[t]{7cm}
\centering
\includegraphics[width=7cm]{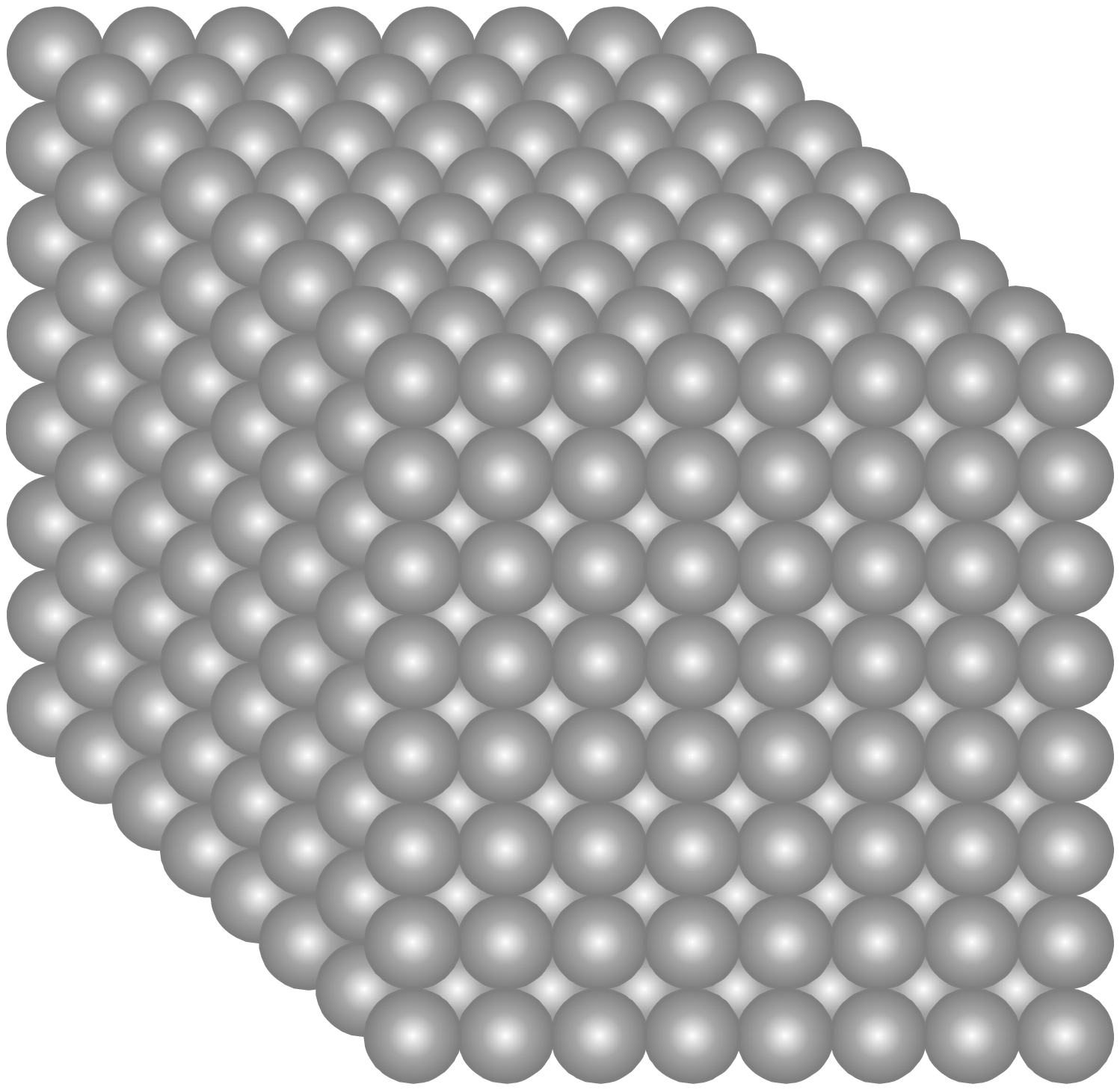}
\caption{\footnotesize Illustration of a space of 8 stacked layers as illustrated in Fig.\ref{Fig.1} of same size, where the interstices are enclosed between three or four stacked balls: the uniformity is only verified within eight orientations.}\label{Fig.2}
\end{minipage}
\end{figure}

\begin{figure}[!h]
\begin{minipage}[t]{7cm}
\centering
\includegraphics[width=7cm]{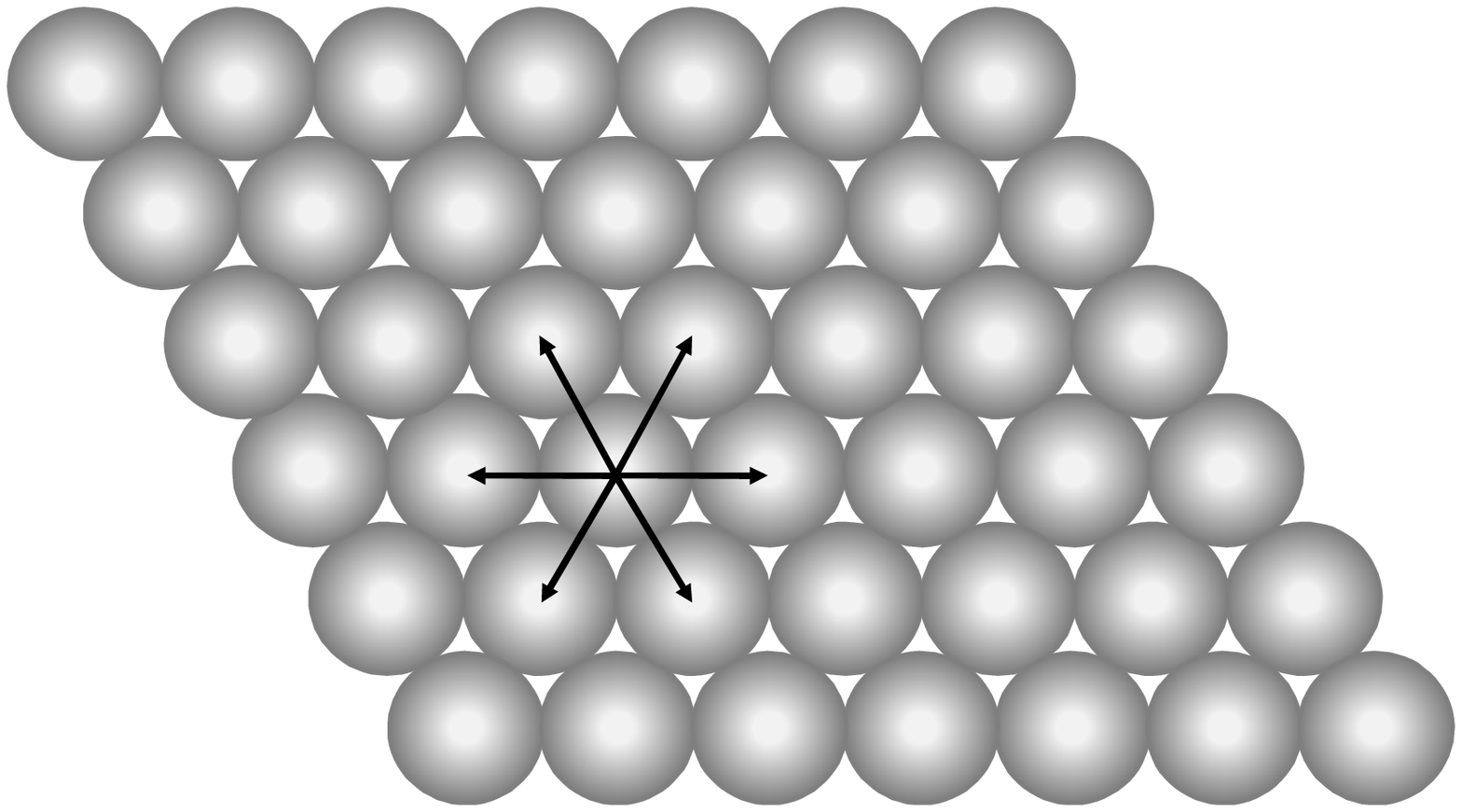}
\caption{\footnotesize Illustration of one layer of $7 \times 6$ equal ball packing, where each interstice is enclosed between three stacked balls: the uniformity is only verified within six orientations.}\label{Fig.01}
\end{minipage}
\hspace*{\fill}
\begin{minipage}[t]{7cm}
\centering
\includegraphics[width=7cm]{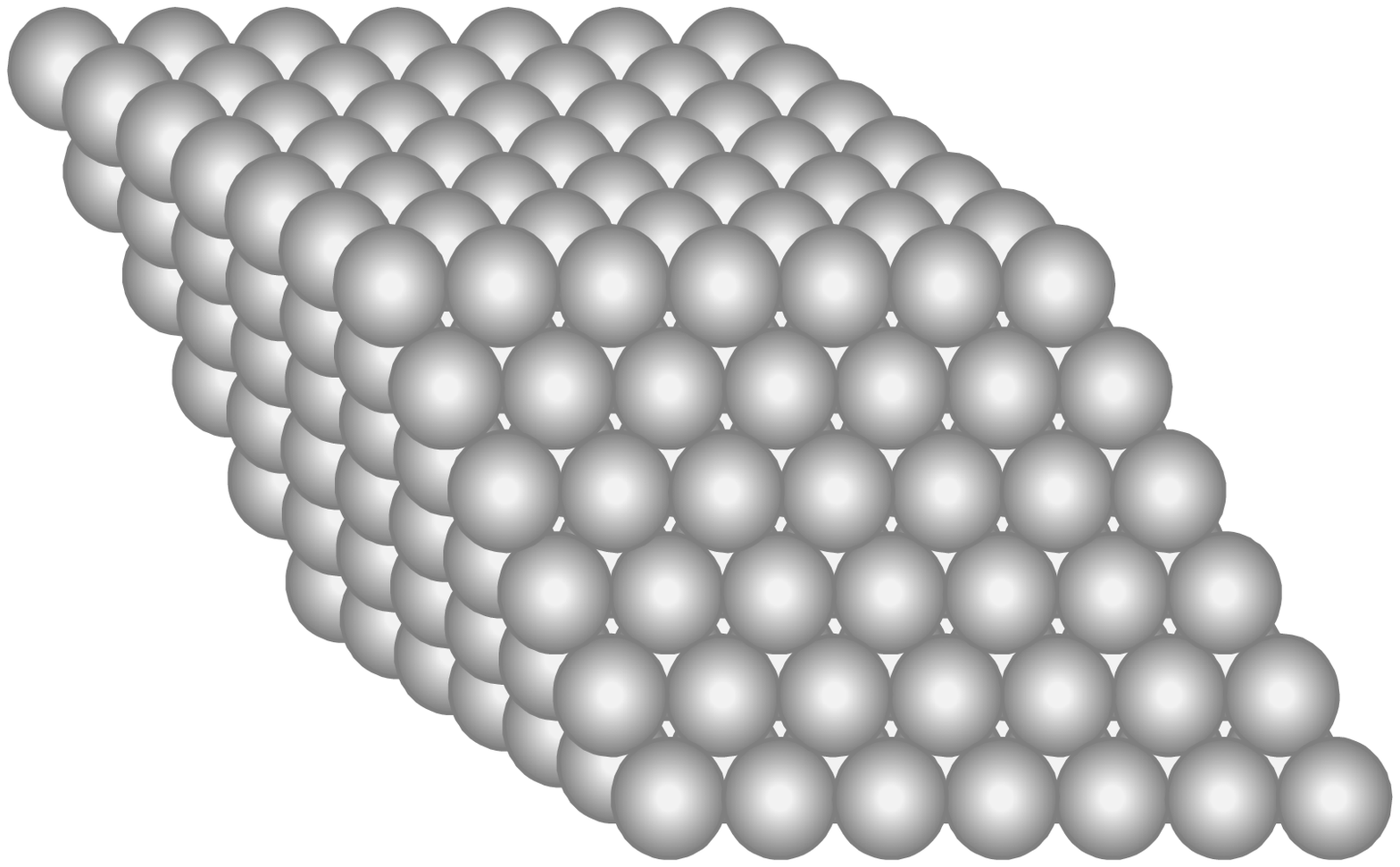}
\caption{\footnotesize Illustration of a space of 6 layers as illustrated in Fig.\ref{Fig.01} of same size, where each interstice is enclosed between three stacked balls: the uniformity is only verified within ten orientations. This packing model is more stable than the one illustrated in Fig.\ref{Fig.2}.}\label{Fig.02}
\end{minipage}
\end{figure}

\begin{figure}[!h]
\begin{minipage}[t]{7cm}
\centering
\includegraphics[width=7cm]{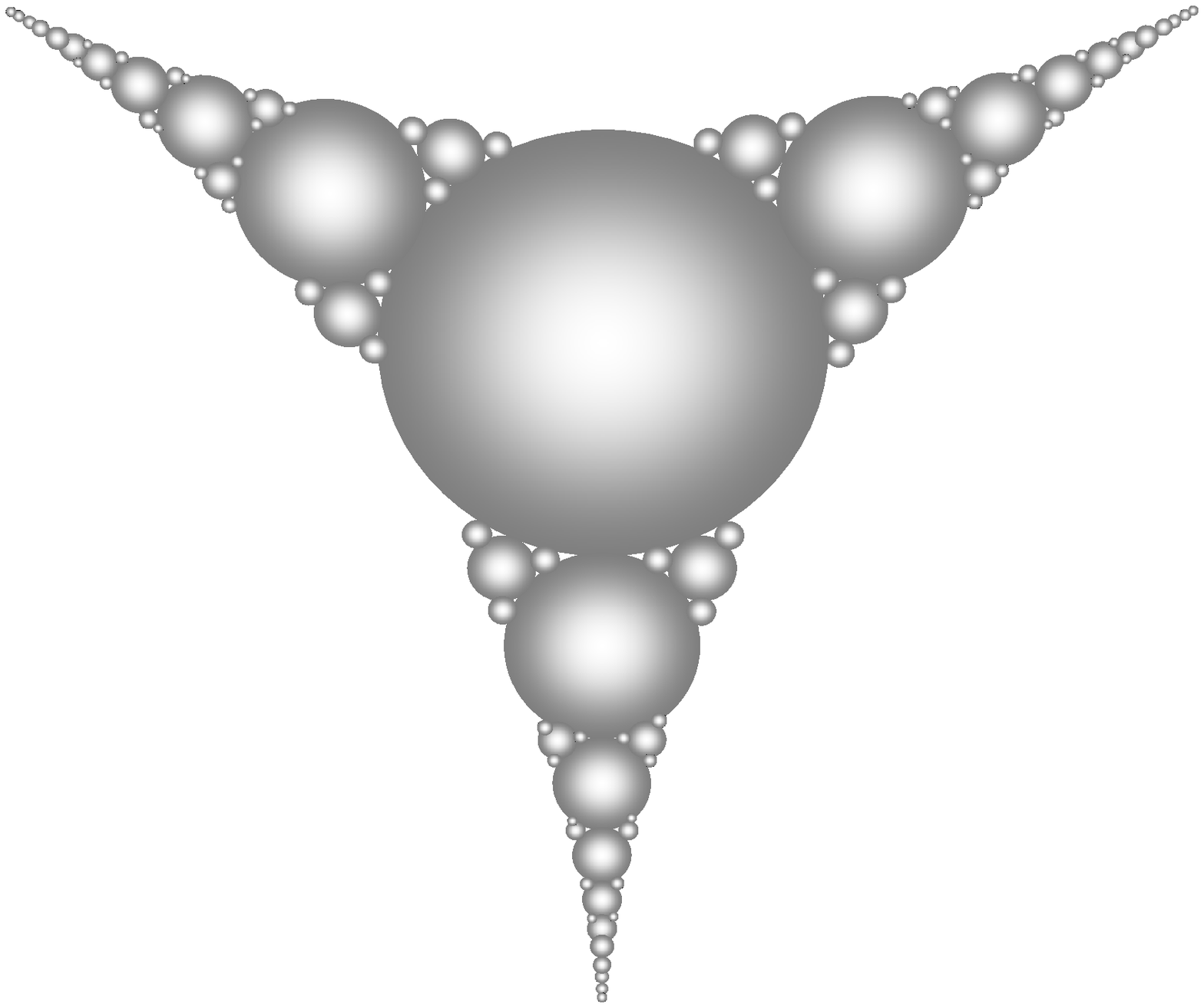}
\caption{\footnotesize Illustration of stacked balls of different sizes to fill the interstice between three stacked equal balls, called Apollonian gasket. The biggest ball is located in the center of the interstice. The smaller the number of stacked equal balls that enclose the interstice is, the more stable the ball packing is.}\label{Fig.3}
\end{minipage}
\hspace*{\fill}
\begin{minipage}[t]{7cm}
\centering
\includegraphics[width=7cm]{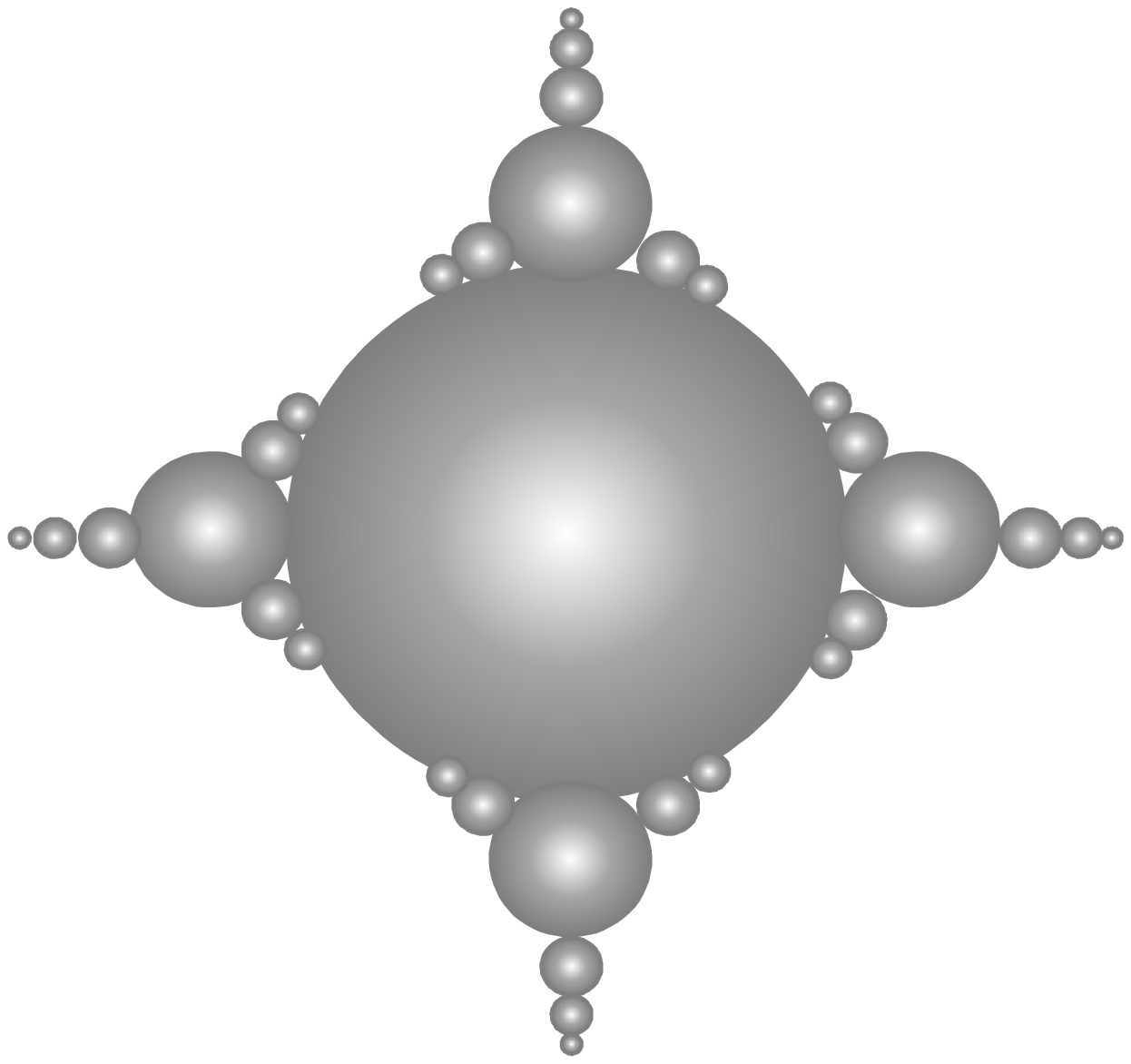}
\caption{\footnotesize Illustration of stacked balls of different sizes to fill the interstice between four stacked equal balls. The biggest ball is located in the center of the interstice.}\label{Fig.4}
\end{minipage}
\end{figure}

An illustration of a ball packing in one layer is given in {Fig.\ref{Fig.1}}, and a volume filled with a ball packing is illustrated in Fig.\ref{Fig.2}, where the interstices are covered by open stacked balls of different sizes to take into account the density of the space-time (as illustrated in Fig.\ref{Fig.3} or in Fig.\ref{Fig.4}). The homogeneity and the expansion can be verified in all directions due to the simultaneously expansion of the stacked open balls. Meanwhile, uniformity is only verified within a limited number of orientations (eight orientations in Fig.\ref{Fig.2}). Indeed, the open ball packing conveys a relative anisotropic expansion referred to certain orientations with different geodesics. A more stable ball packing is illustrated in Fig.\ref{Fig.01} and Fig.\ref{Fig.02} with interstices covered by the Apollonian gaskets of Fig.\ref{Fig.3}. This  ball packing verifies homogeneity and expansion in all directions thanks to the simultaneous expansion of the open balls. However, the uniformity of the space with respect to the geodesics is also verified within limited directions (ten symmetric orientations). Debates about homogeneity and isotropy of the universe can be found in the following selected references \cite{Saad}, \cite{Javan}, \cite{Nath}, \cite{Mig}, \cite{Nada}, \cite{Sylos}, \cite{Yadav}.\pesp

\begin{figure}[!h]
\begin{minipage}[t]{6cm}
\centering
\includegraphics[width=6cm]{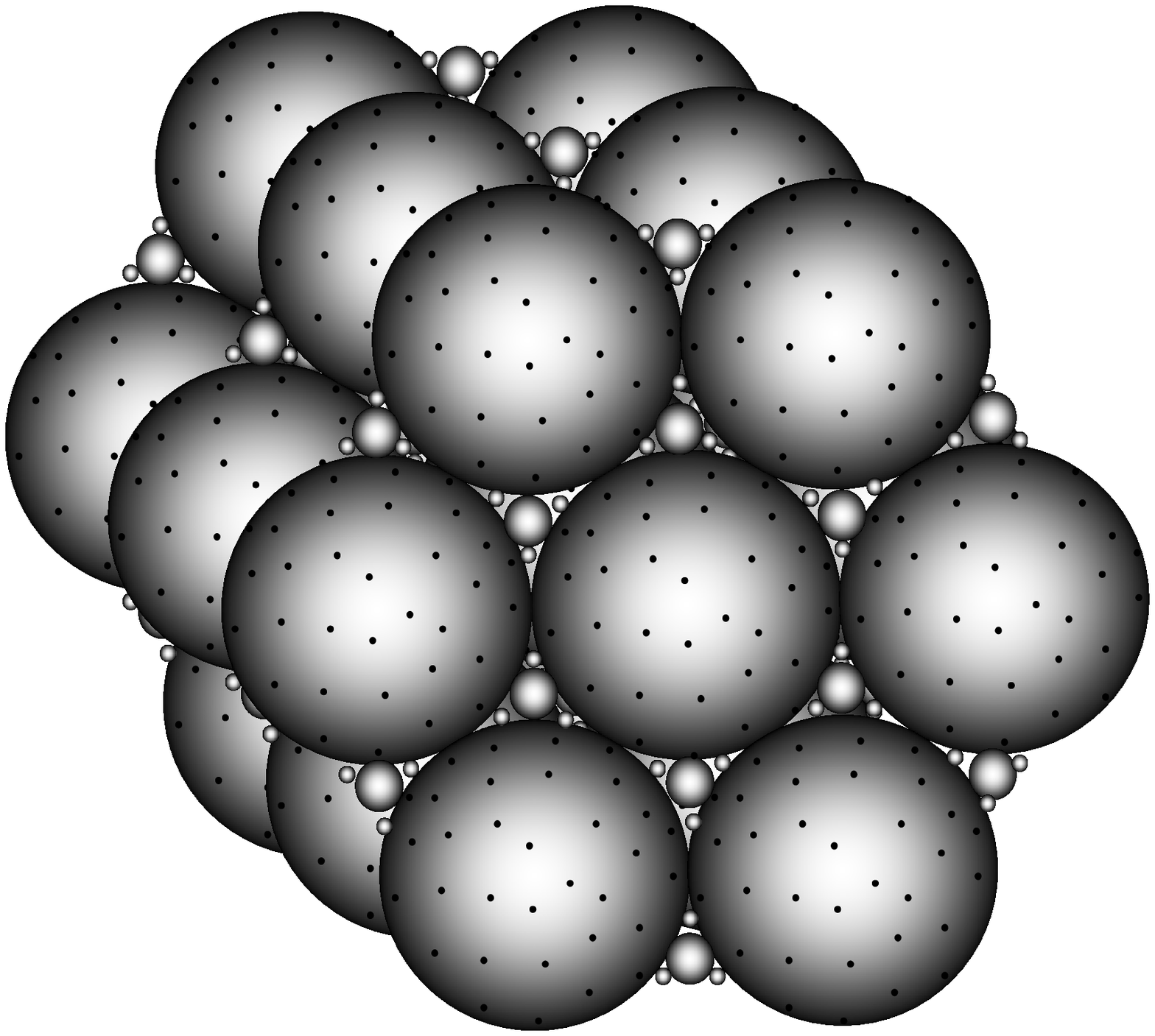}
\caption{\footnotesize Illustration of a portion of stacked equal balls on which matter is represented by dots. Since matter exists within a continuum space-time of four dimensions, then the boundary of each open ball illustrates a reduced form of a bent continuum space-time of four dimensions, boundary of an expanding object of five dimensions. An open ball in the model illustrates a reduced form in dimension of the expanding object of five dimensions. The objects of five dimensions cannot be visualized; nevertheless this model allows to approximate the visualization of the homogeneity and isotropy.}\label{Fig.5}
\end{minipage}
\hspace*{\fill}
\begin{minipage}[t]{6cm}
\centering
\includegraphics[width=6cm]{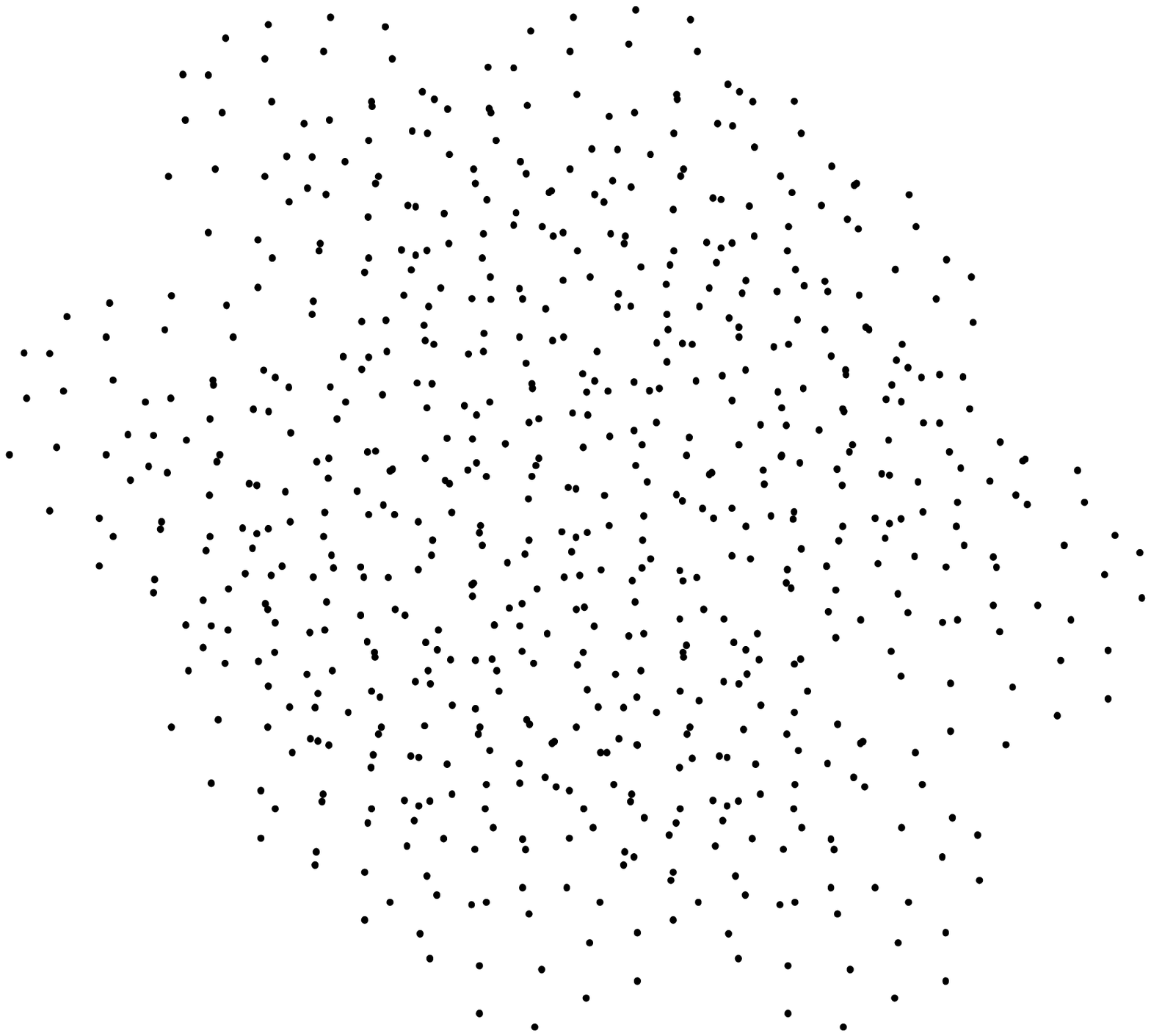}
\caption{\footnotesize The invisibility of the geometry of the stacked equal balls means that the illustration {Fig.\ref{Fig.6}} is the same illustration as {Fig.\ref{Fig.5}} without visualization of the boundaries of the stacked balls. Matter represented by dots appears to be held in the space without pillars (the spheres) that one can see. Homogeneity and isotropy are approximately verified within this model since there is no privileged direction revealed by observation.}\label{Fig.6}
\end{minipage}
\end{figure}

Nevertheless, the non uniformity in the above examples is referred to the non existence of identical geodesics in all directions. However since geodesics are related to the space-time geometry, and the geometry of our universe remains invisible to observation (only matter can be seen), then the consideration of matter distribution within the surface of those expanding stacked open balls may approximate homogeneity to satisfactory needs. An example of illustration of the distribution of matter as dots on the surface of a portion of stacked open balls at large scale conveys the idea of homogeneity, isotropy and expansion. Indeed, the simultaneous expansion of balls in the illustration Fig.\ref{Fig.5} increases the distance between dots in all direction and simulates the space expansion. Meanwhile the observation of dots without geometry (illustrated in {Fig.\ref{Fig.6}}) reflects isotropy and homogeneity of matter distribution regardless of the direction of observation, and simulates the expansion of the space where the dots are plotted. \pesp

The plan of this paper is summarized as follow: in section 2 the change of the state function and entropy generation is introduced, as well as its properties. In section 3 we use asymptotic estimation to define singularities and the space-time of singularities. In section 4 an interpretation of the space-time earliest conditions and gravity origin are presented if matter is considered, including a description of the initial space before the expansion event, the period of the earliest inflation, the period of the space-time of singularities transformation, gravity origin, and the period of primordial space-time when light was released for the first time. In section 5 the conclusion.

\section{Change of the state function and  entropy generation}

It is known that the change in the state function (entropy) can be determined at thermal equilibrium in terms of the changes of system's energy and temperature as a heat (Clausius' entropy \cite{Claus1}), or it can be determined for an isolated system on the total number of micro-states (Boltzmann's entropy \cite{Bolt1}, \cite{Bolt2}), or it can be determined via a probability distribution of the system (Shannon's measure of information \cite{Shann1}, \cite{Shann2}), or in a generalized quantum gravity entropies such as (\cite{Nojiri2}, \cite{Nojiri1}, \cite{Odint}). Nevertheless, the different entropy's definitions mentioned above concern matter characteristics, transformations or distributions,
despite the fact that the universe is made of the space-time (the container) and of all form of matter, energy and radiation (the content).\pesp

To provide a state function sensitive to the space-time transformation consistent with the fundamental principle of cosmology, we use the model of space-time introduced in \cite{BFP}. This model is defined by an accumulation of an infinite number of equal open balls (called basic elements and quantified in the subsection below). The simultaneous expansion of all basic elements of the model simulates the space-time expansion and warranties its homogeneity and isotropy at large scale. Meanwhile, taking into account the density of the space-time, all the interstices are covered by a family of stacked open balls of different sizes known as Apollonian gasket and illustrated in Fig.\ref{Fig.3}. The open balls of different sizes that form the Apollonian gasket within each interstice warranties the non homogeneity and the anisotropy of the space-time at small scale. The initial space of the model is taken to be a limit space as the radii of all balls tend to zero simultaneously. This model simulates the space-time expansion and leads to the following outcomes:\pesp

\begin{itemize}
  \item it leads to understand what happens to the geodesics as the space-time expands via simultaneous expansion of its basic elements. The new result asserts that there exists an infinite number of transversal oscillating paths of least time between any two distant locations, and the straight line geodesics do not exist between any two distant locations of the expanding space-time. The local curvature of each path of least time is equal to $1\over r$, where $r$ is the radius of the expanding equal open balls. As the radii of the stacked open balls increase simultaneously, the space-time expands and the local curvature of each path of least time decreases. The local curvature $1\over r$ of each path of least time corresponds to the normal curvature of each boundary of the expanding open balls. This normal curvature is sensitive to the space-time expansion and will be called the normal curvature of the space-time.

  \item it allows to determine matter distribution at large scale using the infinity of paths of least time and to deduce an optimal shape of the real space-time where matter resides. The model indicates that the distribution of matter is located within the boundaries of the expanding open balls (as illustrated in Fig.\ref{Fig.5}). Since matter exists within a continuum space-time of four dimensions, then the boundary of each open ball illustrates a reduced form of a bent continuum space-time of four dimensions, boundary of an expanding object of five dimensions. The open balls in the model illustrate a reduced form in dimension of the expanding object of five dimensions.

   \item it raises a fundamental distinction between recession movement of matter (the content) and space-time expansion (the recipient), and indicates that the infinite family of stacked equal open balls that simulates in the model the simultaneous expansion of the space-time represents approximatively 74$\%$ of the space-time (Kepler conjecture \cite{Hales}), meanwhile the interstices represents 26$\%$.
\end{itemize}

Moreover it is known that in geometrical optics the interpretation of the interference pattern has failed from its beginning because it is not possible to reproduce interference patterns using the straight line paths. However the use of an infinite number of oscillating paths of least time in phase introduced in \cite{BFP} reproduces the interference patterns observed in the Young's double slit experiment \cite{Young} of 1802, and provides an adequate interpretation of all its characteristics \cite{FBA}. In addition it has also led to a new development about the Van Der Waals torque, as well as a detailed description of the Casimir attraction/repulsion's unknown mechanism in nanotechnology \cite{FBA}.
For the large scale, the model has also led to obtain a compatible interpretation  of the cosmological data of supernovae (SNe) of type Ia \cite{FBA1}, that was the outstanding discovery  of the Nobel laureates of 2011 (\cite{PE},\cite{RI},\cite{RI1}) conveying a shocking reverse of understanding of the state of the universe expansion.\pesp

Thus the compatibility of the model with the recent cosmological data (SNe) for large scale as well as the new development in geometrical optics for the small scale justifies its use to construct a new entropy sensitive to the variation of the normal curvature of the space time. Nevertheless, some notations in relation with the needed formalism from the quantification of the model in \cite{BFP} will be presented below.

\subsection{Quantification adjustment}

The transformations of the space-time are quantified in \cite{BFP} by the sequence ${\mathcal E}=({\mathcal E_n})_{n\geq0}$: the initial space-time is denoted by ${\mathcal E_0}$ and the present space-time is denoted by ${\mathcal E_n}$. If we denote $T$ the age of the expansion of the space-time from the initial space-time ${\mathcal E_0}$ to the present space-time ${\mathcal E_n}$, and $n$ the integer that represents a  subdivision of $T$ into $n$ equal periods of time $T_s$. Minor changes of notation will be taken into account as follow:
\begin{enumerate}
  \item The primordial space-time will be denoted ${\mathcal E_1}$. It will represent in this quantification the surface of the last scattering from where the cosmic microwave background (CMB) comes. The primordial space-time ${\mathcal E_1}$ appears 380 000 years after the Big Bang, when the universe changes from opaque to transparent and releases light to travel for the first time to reveal the cosmic microwave background for observation.

  \item The initial space will be denoted ${\mathcal E_i}$ instead of ${\mathcal E_0}$ to avoid confusion. The initial space ${\mathcal E_i}$ represents an approximation of the space before the Big Bang. The basic elements of the initial space ${\mathcal E_i}$ are the limit of the basic elements of the primordial space-time ${\mathcal E_1}$ as their tiny radii tend to zero simultaneously. The earliest transformations and conditions will be determined using the characteristics of the limit.

  \item The space-time expansion is quantified using a subdivision of the age $T$ of the space-time expansion starting from the initial space ${\mathcal E_i}$ to the present space-time ${\mathcal E_n}$ after $n$ discrete expansions. The period of time $T_s$ represents the elapsed time between any two successive expansions of the space-time. The ratio of $T$ to $T_s$ is the integer $n={T\over T_s}$ that represents the number of subdivisions of the age of the space-time expansion. The bigger the number $n$ of subdivisions of $T$ is, the smaller the period of time $T_s$ between two successive expansions is. Conversely, the smaller the number $n$ of subdivisions of $T$ is, the bigger the period of time $T_s$ between two successive expansions is. Thus the space-time expansion from the primordial space time ${\mathcal E_1}$ to the present space-time ${\mathcal E_n}$ will be quantified by the sequence ${\mathcal E}=({\mathcal E_n})_{n\geq1}$. The earliest conditions of the space-time expansion between the initial space ${\mathcal E_i}$ and the primordial space-time ${\mathcal E_1}$ will be quantified separately.
\end{enumerate}

\subsection{Modeling Adjustment}\label{SE}

In this model the expansion of the space-time is generated by the simultaneous expansion of the stacked open balls of same size. The boundary of each open ball simulates a reduction in dimension of the bent space-time where matter resides. To involve in a state function a variable normal curvature dependent of the expansion of the space-time, the boundaries of the stacked open balls are a good print of the expansion of the space-time. Therefore the boundary of each open ball will be considered rather than the open ball, and we have the following:\pesp

{\bf 1. The initial space ${\mathcal E_i}$} is the space whose basic elements are limit entities with a topology close to that of a point. The basic element of ${\mathcal E_i}$ are defined by the limit as the radius of the basic elements of the primordial space-time ${\mathcal E_1}$ tends to zero.\pesp

{\bf 2. The primordial space-time ${\mathcal E_1}$} is the first space-time accessible for cosmological observation at the step 1. The space-time ${\mathcal E_1}$ is defined by an infinite family of stacked spheres (rather than stacked open balls) with identical tiny radius $r\in]0,1[$, the smallest size of sphere that physics can decern from a dimensionless point. These identical spheres define the basic elements of the primordial space-time ${\mathcal E_1}$.\pesp

{\bf 3. The space-time ${\mathcal E_n}$} is the space-time ${\mathcal E_1}$ after $n$ simultaneous expansion of its basic elements. It is defined by an infinite family of stacked basic elements with identical size, quantified by stacked spheres of same size with radius given by
\begin{equation}\label{RN}
R_{n}={ r\Big(\prod_{j=1}^{n}a_{j-1}\Big)}
\end{equation}
where the step-expanding parameters $(a_n)_{n\geq0}$ form a numerical sequence that quantifies the discrete expansion of the space-time for all $n\geq0$, and satisfies for $a_0=1$ the following conditions:\pesp

i) for all $n\geq1$,
\begin{equation}\label{ai}
a_n>1,
\end{equation}

ii) for all $n\geq1$,
\begin{equation}\label{Dec}
a_{n+1}<a_n,
\end{equation}

iii) the product
\begin{equation}\label{EXP}
\prod_{1}^na_{j-1}\qquad\hbox{is convergent.}
\end{equation}

The normal curvature of each basic element is defined for all  $n\geq1$ by
\begin{equation}\label{CURV}
\delta_n(r)={1\over R_n}={1\over  r(\prod_{j=1}^{n}a_{j-1})}.
\end{equation}

The parameter $a_j$ is called the $j^{th}$ quantified expanding parameter of the space-time between the steps $j$ and $j+1$, and the
product $\prod_{j=1}^{n} a_{j-1}$ is called the expanding parameter of the space-time until the step $n$.\pesp

{\bf 4. The interstices:} the discrete expansion of the space-time ${\mathcal E}=({\mathcal E_n})_{n\geq1}$ simulates an homogeneous, isotropic and expanding space-time that expands via simultaneous expansion of its basic elements. Thus as the stacked equal spheres expand, they create an increasing uncovered narrow space between them called the interstice. To simulate a fully covered dense space by spheres, the interstices are filled with Apollonian gaskets of stacked spheres of different sizes. This process is iterated as the stacked spheres of different sizes expand and create different interstices in turn. Therefore to conserve the density of the space as it expands, all increasing interstices are in turn filled with stacked spheres of different sizes. These repeating imbedded interstices, filled with stacked spheres of different sizes are generated by the space-time expansion and convey the non homogeneity and the anisotropy of the space-time locally.\pesp

\begin{figure}[!h]
\centering
\includegraphics[width=6cm]{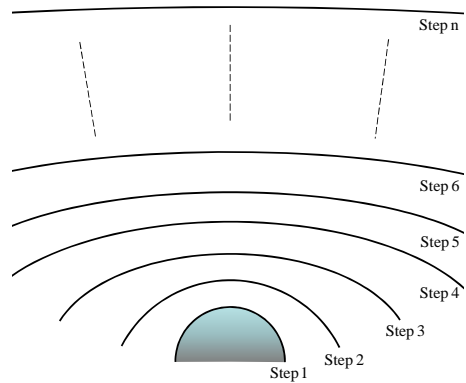}
\caption{\footnotesize Illustration of the expansion of one basic element quantified from step 1 to step $n$. The greater the radius of the basic element is,  the lower the normal curvature of the space-time is.}\label{Fig.8}
\end{figure}

{\bf 5. The normal curvature of the expanding space-time ${\mathcal E}=({\mathcal E_n})_{n\geq1}$} is approximatively determined by the normal curvature of the expanding basic elements of the space-time. Indeed, since all basic elements are identical, then the normal curvature of the space-time is given by the normal curvature of one basic element which is a sphere with radius defined by (\ref{RN}) for all $n\geq1$.
Thus since all paths of least time on a sphere are great circle arcs of radius (\ref{RN}), then the normal curvature of a sphere is given by the normal curvature (\ref{CURV}) of the great circle arc for all $n\geq1$. The greater the radius of the basic element is, the lower the normal curvature of the space-time is (as illustrated in Fig.\ref{Fig.8}).\pesp

The normal curvature of the spheres in the interstices related to the non homogeneity and the anisotropy of the space-time is not considered since it depends on the normal curvature of the basic elements of the space-time. Therefore, the definition of the space-time state function will be related only to the normal curvature of the expanding basic elements of the space-time that provides homogeneity, isotropy at large scale and generates the space-time expansion.

\subsection{Space-time state function}

Using the normal curvature (\ref{CURV}) of the stacked spheres of the expanding space-time ${\mathcal E}=({\mathcal E_n})_{n\geq1}$, we define the space-time state function as follow:

\begin{defn}
Consider the space-time $\mathcal E=({\mathcal E_n})_{n\geq1}$ where ${\mathcal E_1}$ is the primordial space-time. For an arbitrary tiny radius $r\in]0,1[$ and an arbitrary finite positive constant k, we define the entropy of space-time $\mathcal E$ at the Step n for all $n\geq1$ as
\begin{equation}\label{ENT}
S_n(r)=-k\ \ln (\delta_{n}(r)),
\end{equation}
where $\delta_n(r)={1\over R_{n}}$  is the normal curvature of the basic elements of radius $R_{n}={ r\prod_{1}^{n}a_{j-1}}$ of the space-time ${\mathcal E_n}$ at the step $n$.
\end{defn}

The function (\ref{ENT}) is a state function because it is given by an integral that depends only on the lower limit 1 and the upper limit $\delta_n(r)$. The lower limit is given by the normal curvature of a unit sphere. The upper limit is given by the normal curvature of an expanding basic element of the space-time ${\mathcal E_n}$, quantified by the radius given by (\ref{RN}) for all $n\geq1$. Indeed, the function (\ref{ENT}) can be written as
\begin{equation}
S_n(r)=-k\int_1^{\delta_{n}(r)}{1\over t}\ dt=-k\ \Big(\ln (\delta_{n}(r))-\ln (1)\Big)=-k\ \ln (\delta_{n}(r)).
\end{equation}

\subsection{Quantified entropy generation and entropy generation}

Using the above state function given by (\ref{ENT}), the quantified change of the state function together with the space-time expansion, denoted by $\Delta S_n$, is given by
\begin{equation}\label{Delt}
\Delta S_n=S_{n}(r)-S_{n-1}(r)
\end{equation}
where $S_n(r)$ is the state function of the space-time $\mathcal E_n$ at the Step $n$, and $S_{n-1}(r)$ is the state function of the space-time $\mathcal E_{n-1}$ at the Step $n-1$ for all $n>1$.\pesp

\ni{\bf Notation}\pesp

a) We call{\it "quantified entropy generation between the steps $n-1$ and $n$"} of the space-time $\mathcal E=({\mathcal E_n})_{n\geq1}$ and we denote it by $\Delta S_n=S_{n}(r)-S_{n-1}(r)$, the quantified change of the state function of successive expansions of the space-time $\mathcal E$ for all $n>1$.\pesp

b) We call {\it "space-time entropy generation between the steps 1 and n"} of the space-time $\mathcal E=({\mathcal E_n})_{n\geq1}$ and we denote it by $\Delta \mathbb{S}_n=S_{n}(r)-S_1(r)$, the change of the state function between the primordial space-time $\mathcal E_1$  (at the Step 1) and the present space-time $\mathcal E_n$ (at the Step $n$ for all $n>1$).\pesp

Thus for all $n>1$ the quantified entropy generation between the steps $n-1$ and $n$ of the space-time  $\mathcal E=({\mathcal E_n})_{n\geq1}$  is given by
\begin{equation}\label{statef}
\Delta S_n=k\ \ln({\delta_{n-1}(r)\over \delta_{n}(r)})
\end{equation}
 where $\delta_n(r)={1\over R_{n}}$ is the normal curvature of the basic element of the space-time $\mathcal E_n$ at the Step $n$, and using (\ref{statef}) we have
\begin{equation}
S_{n}(r)=S_{n-1}(r)+k\ \ln({\delta_{n-1}(r)\over \delta_{n}(r)}).
\end{equation}

\begin{rmak}
i) The arbitrary positive constant $k$  in the equality (\ref{statef}) can be considered as the Boltzmann constant if needed. The state function (\ref{ENT}) is defined by an increasing function of $\delta_{n}(r)$. Indeed, using (\ref{CURV}) the normal curvature of the space-time ${\mathcal E_n}$ for all $n>1$ is a decreasing function of $n$ (for an arbitrary constant $r\in]0,1[$), and we have
\begin{equation}
\forall n>1, \qquad \delta_{n}(r)<\delta_{n-1}(r),
\end{equation}
which makes (\ref{ENT}) an increasing function of $\delta_{n}(r)$ as $n$ increases.\pesp

ii)  The quantified entropy generation (\ref{statef}) between the steps $n-1$ and $n$ can be written using integral notation as follow
\begin{equation}
\Delta S_n=k\ln \Big({\delta_{n-1}(r)\over \delta_{n}(r)}\Big)=-k\ \int_{\delta_{n-1}(r)}^{\delta_{n}(r)} {1\over t}\ dt.
\end{equation}
\end{rmak}

\subsection{Properties of the space-time state function}

It is known that the normal curvature measures the local degree of deformation of a curve (or a surface). It is a quantity that measures the speed at which the graph of a given curve (or surface) deviates from the direction of the tangent line in the neighborhood of local points. Since the basic elements of the space-time $\mathcal E_n$ are expanding spheres, then the equality (\ref{Delt}) for all $n>1$ verifies the following:

\begin{prp} Let $(S_{n})_{n\geq1}$ be the sequence of state functions of the space-time ${\mathcal E}=(\mathcal E_n)_{n\geq1}$, then the entropy generation between the steps 1 and n of the space-time ${\mathcal E}$ verifies for all $n>1$ and $r\in]0,1[$
\begin{equation}\label{SN1}
\Delta \mathbb{S}_n=S_{n}(r)-S_1(r)=k\ln({\delta_1(r)\over\delta_{n}(r)}).
\end{equation}
\end{prp}

{\it Proof}: Using the space-time entropy generation between successive steps (\ref{Delt}) for all $n>1$, we obtain
\begin{equation}\label{eqn}
\begin{array}{cc}
  \Delta S_{n} =& S_{n}(r)-S_{n-1}(r) \\
\Delta S_{n-1}= & S_{n-1}(r)-S_{n-2}(r) \\
\vdots\quad \vdots & \vdots \\
  \Delta S_2= & S_{2}(r)-S_1(r)
\end{array}
\end{equation}
by adding the $(n-1)$ equalities (\ref{eqn}), it gives
\begin{equation}\label{Some}
  \Delta S_{n}+\Delta S_{n-1}+\ldots+\Delta S_2= S_{n}(r)-S_1(r)
\end{equation}
and using (\ref{statef}), the above equality is simplified to
\begin{equation}
S_{n}(r)=S_1(r)+k\ln({\delta_1(r)\over\delta_{n}(r)}).
\end{equation}
The result can also be obtained directly by substituting (\ref{ENT}) in the definition of $\Delta \mathbb{S}_n$.

\begin{prp}
The quantified entropy generation  (\ref{statef}) between the steps $n-1$ and $n$ of the space-time ${\mathcal E}=({\mathcal E_n})_{n\geq1}$ verifies
\begin{equation}\label{Cst}
(i)\ \forall n>1,  \quad  \Delta S_n=0\quad \Longleftrightarrow \quad {\mathcal E}=({\mathcal E_n})_{n\geq1}\quad \hbox{is a static space-time},
\end{equation}
\begin{equation}\label{Expand}
(ii)\ \forall n>1,  \quad  \Delta S_n>0\quad \Longleftrightarrow\quad {\mathcal E}=({\mathcal E_n})_{n\geq1} \quad\hbox{is an expanding space-time},
\end{equation}
\begin{equation}\label{Contract}
(iii)\ \forall n>1,  \quad  \Delta S_n<0\quad \Longleftrightarrow\quad {\mathcal E}=({\mathcal E_n})_{n\geq1} \quad\hbox{is a contracting space-time}.
\end{equation}
\end{prp}

{\it Proof}: (i) For all $n>1$, we have $\Delta S_n=0$, then using (\ref{statef}) we have $k\ln \Big({\delta_{n-1}(r)\over \delta_{n}(r)}\Big)=0$, which gives using the normal curvature (\ref{CURV})
\begin{equation}
k\ln \Big({R_{n}\over R_{n-1}}\Big)=0,
\end{equation}
 that is to say for all $n>1$, $R_{n}=R_{n-1}$, which characterizes a static space-time. The converse is evident.

(ii) For all $n>1$, we have $\Delta S_n>0$, then using (\ref{statef}), it gives $k\ln ({R_{n}\over R_{n-1}})>0$, which gives, for all $n>1$, $\ln (R_{n})>\ln (R_{n-1})$. That is to say, for all $n>1$, $R_{n}>R_{n-1}$, which characterizes an expanding space-time. The converse is evident.

iii) For all $n>1$, we have $\Delta S_n<0$, that gives $-k\ln ({R_{n}\over R_{n-1}})>0$, which means that  $k\ln ({R_{n-1}\over R_{n}})>0$. Then, for all $n>1$, $\ln (R_{n-1})>\ln (R_{n})$. That is to say, for all $n>1$, $R_{n-1}>R_{n}$, which characterizes a contracting space-time. The converse is evident.\pesp

\begin{prp}
The quantified entropy generation between the steps $n-1$ and $n$ of the space-time ${\mathcal E}=({\mathcal E_n})_{n\geq1}$  measures the space-time quantified expansion. In particular for all $n>1$
\begin{equation}\label{an1}
 \Delta S_n=k\ln (a_{n-1})
\end{equation}
with $a_{n-1}$ the $(n-1)^{th}$ space-time expanding parameter from the Step $n-1$ to the Step $n$.
\end{prp}

{\it Proof}: The substitution of the normal curvature (\ref{CURV}) in the entropy generation (\ref{statef}) gives
\begin{equation}\label{i+1}
\Delta S_n=k\ \ln\Big({\delta_{n-1}(r)\over \delta_{n}(r)}\Big)=k\ln\Big({R_{n}\over R_{n-1}}\Big)=k\ln\Big({r\Big(\prod_{i=1}^{n}a_{i-1}\Big)\over r\Big(\prod_{i=1}^{n-1}a_{i-1}\Big) }\Big)=k\ln( a_{n-1}),
\end{equation}
which completes the proof.

\begin{prop}\label{Prop1}

i) For $n>1$ the quantified entropy generation between the steps $n-1$ and $n$ of the space-time ${\mathcal E}=({\mathcal E_n})_{n\geq1}$ for a short period of time between successive steps of expansion verifies
\begin{equation}\label{DS0}
\Delta S_n\approx 0
\end{equation}

ii) For $n>1$ the quantified entropy generation between the steps $n-1$ and $n$ of the space-time ${\mathcal E}=({\mathcal E_n})_{n\geq1}$ for a large period of time between successive steps of expansion verifies
\begin{equation}\label{Ap1}
\Delta S_n>0
\end{equation}
\end{prop}

{\it Proof}: i)  The expanding parameter (\ref{EXP}) of the space-time ${\mathcal E}=({\mathcal E_n})_{n\geq1}$ can be written  as
\begin{equation}
\prod_{i=1}^{n} a_{i-1}=e^{\sum_{1}^{n} \ln (a_{i-1})},
\end{equation}
and as $n$ tends to infinity, the sequence (\ref{EXP}) has to be convergent. Therefore there exists a big $N>0$ such that for all $n-1\geq N$, $\ln (a_{n-1})$ tends to 0, which gives for all $n-1\geq N$, the parameter $a_{n-1}$ tends to 1, then we have for all $n-1\geq N$, the parameter $a_{n-1}\approx1$ that gives using (\ref{an1})
\begin{equation}
 \Delta S_n=k\ln (a_{n-1})\approx0.
\end{equation}
Since $n-1\geq N$, then $n$ is a big integer and represents the number of subdivisions of the age $T$ of the expanding space-time ${\mathcal E}$ at the step $n$. Therefore the bigger the number $n$ is, the shorter the period of time interval between two successive expansions is. For this short period of time, the quantified entropy generation (\ref{statef}) verifies $\Delta S_n\approx 0$.\pesp

ii) For all $n-1 <N$, the integer $n$ is less than $N+1$ a big number and represents the number of subdivisions of the age $T$ of the expanding space-time ${\mathcal E}$ at the step $n$. Therefore, the smaller the number $n$ is, the larger the period of time interval between successive expansions is. Thus for a large period of time interval between successive expansions such that $n-1<N$, the $(n-1)^{th}$ quantified expanding parameter $a_{n-1}$ of the space-time ${\mathcal E}$ verifies (\ref{ai}), that is to say for a small integer $1< n-1<N$, we have $a_{n-1}>1$. Since $k$ is a positive number, then we have
\begin{equation}
 k\ln (a_{n-1})>k\ln (1),
\end{equation}
and using (\ref{an1}), the inequality gives
\begin{equation}
 \Delta S_n>0
\end{equation}
for a large period of time interval between successive expansions.

\begin{prop}\label{P2}
If $(S_{n})_{n\geq1}$ is the sequence of state functions of the space-time ${\mathcal E}=({\mathcal E}_n)_{n\geq 1}$, then for all $n\geq1$,

i) the entropy generation between the steps 1 and n of the space-time ${\mathcal E}$ due to its expansion from the primordial space-time ${\mathcal E_1}$ to the space-time ${\mathcal E_n}$ is increasing as $n$ increases, and we have
\begin{equation}\label{EG}
\Delta\mathbb{S}_n=S_n(r)-S_1(r)=k\ \ln ({\prod_{j=1}^{n}a_{j-1} }),
\end{equation}
where $a_{j-1}$ is the $(j-1)^{th}$ quantified expanding parameter of the space-time.

ii) the quantified entropy generation between the steps $n-1$ and $n$ of the space-time ${\mathcal E}$ due to a successive expansion is decreasing as $n$ increases and we have for all $n>1$
\begin{equation}
 \Delta S_{n+1}<\Delta S_n.
\end{equation}

\end{prop}

{\it Proof}: i) Using the state function (\ref{ENT}) and the normal curvature (\ref{CURV}), we have for all $n>0$
\begin{equation}
S_n(r)=-k\ \ln \Big(\delta_{n}(r)\Big)=-k\ \ln \Big({1\over r(\prod_{1}^{n}a_{j-1}) }\Big)= k\ \ln \Big({r(\prod_{1}^{n}a_{j-1}) }\Big)
\end{equation}
then
\begin{equation}\label{Prodai}
S_n(r)= k\ \ln (r)+k\ \ln \Big({\prod_{1}^{n}a_{j-1} }\Big)
\end{equation}
$$= k\ \ln (r)+k\ \sum_{j=1}^n\ln (a_{j-1}).$$

Using (\ref{ENT}) for $n=1$, it gives $S_1(r)=k\ \ln (r)$, then we have
\begin{equation}\label{Eq0}
S_n(r)= S_1(r)+k\ \ln ({\prod_{j=1}^{n}a_{j-1} }).
\end{equation}

since $\Delta\mathbb{S}_n=S_n(r)-S_1(r)$ for all $n>1$, then we have (\ref{EG}), which gives
\begin{equation}\label{EnG}
\Delta\mathbb{S}_n=S_n(r)-S_1(r)=k\ \sum_{j=1}^n\ln (a_{j-1}).
\end{equation}

Using (\ref{ai}), we have for all $j-1>0$, $a_{j-1}>1$, which gives $\ln(a_{j-1})>0$, therefore the entropy generation (\ref{EnG}) is given by an increasing convergent series with positive terms. Thus the entropy generation (\ref{EnG}) of the space-time ${\mathcal E}$ is increasing as $n$ increases.\pesp

ii) From (\ref{an1}) we have for $i>0$
\begin{equation}\label{aj}
 \Delta S_i=k\ln (a_{i-1}),
\end{equation}
and the property (\ref{Dec}) gives for all $i>0$
\begin{equation}
a_{i}<a_{i-1},
\end{equation}
then
\begin{equation}
\ln (a_{i})<\ln (a_{i-1}),
\end{equation}
and using the finite positive constant $k>0$ it gives
\begin{equation}\label{ln}
k\ln (a_{i})<k\ln (a_{i-1}).
\end{equation}

The use of (\ref{aj}) (for index $i$ and $i+1$) in the inequality (\ref{ln}) gives for all $i>0$
\begin{equation}\label{SI}
 \Delta S_{i+1}<\Delta S_i,
\end{equation}
that is to say the quantified entropy generation $\Delta S_n$ is decreasing as $n$ increases.

\begin{rmak}\label{IMP}
i) The approximation (\ref{DS0}) means that the variation of space-time normal curvature remains imperceptible for a short period of time $T_s$ between two successive space-time ${\mathcal E_i}$ and ${\mathcal E_{i+1}}$ for all $i\geq1$. A short period of time $T_s$ is related to the age $T$ of the space-time expansion at the step n by the equality $T=n\times T_s$. Since $T_s={T\over n}$, then the period $T_s$ between two successive expansions is short when n is a big integer. Meanwhile the period $T_s$ between two successive expansions is large when n is a small positive integer, and in this case the quantified entropy generation (\ref{Ap1}) is strictly positive.

ii) The quantified entropy generation $\Delta S_n$ between the steps $n-1$ and $n$ for all $n>1$ of the space-time ${\mathcal E}=({\mathcal E_n})_{n\geq1}$ is decreasing as $n$ increases (Proposition \ref{P2}, ii)), meanwhile the entropy generation $\Delta\mathbb{S}_n$ of the space-time between the steps 1 and $n$ of the space-time ${\mathcal E}=({\mathcal E_n})_{n\geq1}$ is always increasing (Proposition \ref{P2}, i)), with a decelerating increase since $\Delta\mathbb{S}_n=\sum_{i=2}^n \Delta S_i$, and $\Delta S_i$ verifies (\ref{SI}) .
\end{rmak}

\section{Asymptotic estimation, Space of asymptotic basic elements}

The space-time expansion from the primordial space-time ${\mathcal E_1}$ to the space-time ${\mathcal E_n}$ is quantified by the sequence of the expanding space-time ${\mathcal E}=({\mathcal E_n})_{n\geq1}$. However, the earliest period of expansion lasted for approximatively 380 000 years after the Big Bang represents in this model the period of expansion between the initial space ${\mathcal E_i}$ and the primordial space-time ${\mathcal E_1}$. An intermediary space-time that quantifies this period can be found to understand the earliest transformation of the initial space ${\mathcal E_i}$. To cope with this unreachable period, the key is conveyed by a limit approach that allows to build characteristics and properties for better understanding. Indeed, from equality (\ref{CURV}) we know that the normal curvature of the space-time ${\mathcal E_1}$ is given for a tiny radius $r\in]0,1[$ by
\begin{equation}\label{DR}
\delta(r)={1\over r}.
\end{equation}

Then as all radii $r$ of the basic elements of the space-time ${\mathcal E_1}$ tend to zero simultaneously, the normal curvature $\delta(r)$ of each basic element of the space-time ${\mathcal E_1}$ tends to the limit
\begin{equation}
\lim_{r\longrightarrow 0}\delta(r)=\lim_{r\longrightarrow 0}{1\over r}=+\infty.
\end{equation}

 From subsection $\S$ \ref{SE} the initial space ${\mathcal E_i}$ represents an approximation of the space before the Big Bang. The basic elements of the initial space ${\mathcal E_i}$ are the limit of the basic elements of the primordial space-time ${\mathcal E_1}$ as their tiny radii tend to zero simultaneously. The initial space ${\mathcal E_i}$ is then characterized by a normal curvature close to plus infinity, and this characteristic leads to others, but first we need to build the following:

\subsection{Set of asymptotic expanding basic spheres}

The divergence of the normal curvature $\delta(r)$ to plus infinity as the radius $r$ tends to zero, means that for all $B>0$ a big real number, there exists a small radius $r_0>0$, such that for all $r\in]0,r_0[$, the normal curvature $\delta(r)>B$. This property leads to introduce a set of asymptotic expanding spheres, denoted by $\mathcal{B}_h(c)$, defined by the disjoint union of spheres $\mathbb{S}(c,r)$ with same center $c$, radius $r\in]0,r_0[$ and a normal curvature close to plus infinity, as follow:
\begin{equation}\label{singularity}
\mathcal{B}_h(c)=\bigcup_{0<r<r_0}\mathbb{S}(c,r).
\end{equation}

The set of expanding asymptotic spheres $\mathcal{B}_h(c)$ represents all continuous transformations of any limit basic element $c$ of the initial space ${\mathcal E_{i}}$ for all $r\in]0,r_0[$. The normal curvature of each transformed sphere $\mathbb{S}(c,r)$ of $\mathcal{B}_h(c)$ remains close to plus infinity. The set $\mathcal{B}_h(c)$ is an open ball without center, defined by a disjoint union of imbedded spheres of radius $0<r<r_0$. Each sphere is the expansion of previous spheres of smaller radii. All spheres of $\mathcal{B}_h(c)$ are homotopic due to the continuous transformation by expansion. The upper limit $r_0$ of the radius $r$ of $\mathbb{S}(c,r)$  depends on the choice of the big constant $B$.

\subsection{Asymptotic expanding space-time ${\mathcal E_{r}}$}

The spheres from the set $\mathcal{B}_h(c)$, with equal radius $r$ and for all center $c$ basic element of the initial space ${\mathcal E_{i}}$, can be used to define an asymptotic expanding space-time. Provided that it incorporates all the transformations of each basic element of the initial space ${\mathcal E_{i}}$ with a topology close to that of a point, to a basic element of the primordial space-time ${\mathcal E_{1}}$ with a topology of a sphere. Indeed, we define this asymptotic expanding space-time as follow:

\begin{defn}
We call  asymptotic expanding space-time and we denote ${\mathcal E_{r}}$ for $r\in]0,r_0[$, the space-time defined by the infinite family of accumulated spheres $\mathbb{S}(c,r)$ of $\mathcal{B}_h(c)$, with equal radius $r$ for all basic element $c\in {\mathcal E_{i}}$.
 \end{defn}

\begin{prp}\label{ProSing}
The asymptotic expanding space-time ${\mathcal E_{r}}$, for all $r\in]0,r_0[$, verifies the following properties:
\begin{enumerate}
  \item The normal curvature of the asymptotic expanding space-time  ${\mathcal E_{r}}$, for all $r\in]0,r_0[$, is close to plus infinity.
  \item The entropy of the asymptotic expanding space-time ${\mathcal E_{r}}$, for all $r\in]0,r_0[$, is close to minus infinity.

  \item The change of the state function due to a continuous expansion of an asymptotic space-time ${\mathcal E_{r_1}}$ to an asymptotic space-time ${\mathcal E_{r_2}}$ for all $0<r_1<r_2<r_0$ is an indeterminate form.

 \item The entropy generation of the continuous expansion of an asymptotic expanding space-time ${\mathcal E_{r}}$, for all $r\in]0,r_0[$, to the primordial space-time ${\mathcal E_{1}}$ is close to plus infinity.

\end{enumerate}
\end{prp}

{\it Proof:} 1. The normal curvature of the asymptotic expanding space-time  ${\mathcal E_{r}}$ is close to plus infinity since it is given by the normal curvature of identical spheres of the open sets $\mathcal{B}_h(c)$ defined by (\ref{singularity}) for all initial basic element $c\in {\mathcal E_{i}}$.\pesp

2. From the definition of entropy (\ref{ENT}), the entropy $S_{h}$ of the  asymptotic expanding space-time ${\mathcal E_{r}}$ is close to minus infinity. Indeed, for all $r<r_0$, $\delta(r)={1\over r}$  tends to plus infinity, which gives
\begin{equation}\label{S0}
S_{h}=\lim_{\delta(r)\longrightarrow +\infty}-k\ln(\delta(r))=-\infty.
\end{equation}

3. To evaluate the entropy generation of the expansion of the  asymptotic expanding space-time ${\mathcal E_{r_1}}$ to the asymptotic expanding space-time ${\mathcal E_{r_2}}$ for all $0<r_1<r_2<r_0$, we use the difference of their state functions denoted respectively $S_{h_1}$ and $S_{h_2}$. Indeed, since $S_{h_1}$ and $S_{h_2}$ are given by (\ref{S0}) for $r=r_1$, respectively $r=r_2$, and since for all $r<r_0$, $\delta(r)={1\over r}$ tends to plus infinity, then we have
\begin{equation}\label{BBANG}
S_{h_2}-S_{h_1}=(-\infty)-(-\infty),
\end{equation}
 which is an indeterminate form.\pesp

 4. To evaluate the entropy generation of the expansion of the asymptotic space-time ${\mathcal E_{r}}$, for all $r\in]0,r_0[$, to the primordial space-time ${\mathcal E_{1}}$, we use the difference of their state functions  $S_{1}(r)-S_{h}$. Indeed, since $S_{1}(r)$ is negative and finite for all $r_0\leq r\ll1$, and $S_{h}$ is given by (\ref{S0}), then we have
\begin{equation}\label{BBANG}
S_{1}(r)-S_{h}=S_{1}(r)-(-\infty)=+\infty,
\end{equation}
 which means that the entropy generation after continuous expansion of the asymptotic expanding space-time ${\mathcal E_{r}}$, for $r\in]0,r_0[$, to the primordial space-time ${\mathcal E_1}$ is close to $+\infty$.
\pesp

\begin{rmak} i) In Property \ref{ProSing}, 3., the entropy generation is found to be an indeterminate form, because the above entropies $S_{h_2}$ and $S_{h_1}$ approach minus infinity, without being equal to minus infinity. That is why the changes of the state function between two distinct asymptotic expanding spaces-time ${\mathcal E_r}$ is not zero, but unknown.

ii) The characteristics introduced in Property \ref{ProSing} convey divergence before the appearance of the primordial space-time ${\mathcal E_1}$. This leads to define a singularity, and a space of singularities.

iii) For all $r\in]0,r_0[$, the surface area to the volume ratio ${A\over V}=3\delta(r)$ of each basic element of the asymptotic expanding space-time ${\mathcal E_r}$ is close to plus infinity.
\end{rmak}

\subsection{Singularity, space of singularities}

Defining singularities when the geometry becomes unreachable is related to the used theory and formalism of application. Indeed, singularities are seen as an end of the space-time, a pathological set of points where our tools are not defined. In mathematics, a singularity of a function refers to a point where the function is not defined. In physics, singularities appear unavoidable and carry a potential of great interest. In cosmology, a singularity refers to collapse of matter in which the laws of physics become indistinguishable. A gravitational singularity is a space location where the gravitational field becomes infinite. In this framework singularities carry a potential of great interest and are defined as follow:

\begin{defn}\label{sing}
We call singularity each sphere of $\mathcal{B}_h(c)$, for all basic element $c$ of the initial space-time ${\mathcal E_{i}}$, with a normal curvature close to plus infinity and a state function, defined by
\begin{equation}\label{SR}
S_r(\delta(r))=-k\ \ln (\delta(r)),
\end{equation}
close to minus infinity, with $k$ an arbitrary positif constant and $\delta(r)$ the normal curvature given by (\ref{DR}).
\end{defn}

\begin{rmak}\label{Remark}
 i) A sphere's normal curvature $\delta(r)={1\over r}$ close to plus infinity, means that for all $B>0$ there exists $r_0>0$ such that for all $r\in]0,r_0[$, the sphere normal curvature $\delta(r)={1\over r}$ is close to plus infinity.

ii) The open set $\mathcal{B}_h(c)$ given by (\ref{singularity}) represents an archive of the continuous transformation of one singularity (all imbedded spheres are homotopic). Indeed $\mathcal{B}_h (c)$ is an open ball without center, defined by a disjoint union of imbedded spheres of radius $0<r<r_0$. Each sphere is a singularity characterized by a normal curvature close to plus infinity and an entropy close to minus infinity.
\end{rmak}

\begin{defn}
We call space of singularities the space defined by an infinite accumulation of identical singularities.
\end{defn}

\begin{rmak}\label{RMAK}
i) Since the basic elements of the asymptotic expanding space-time ${\mathcal E_{r}}$, for all radius $0<r<r_0$, are subsets of $\mathcal{B}_h$ with a normal curvature close to plus infinity, and a state function (\ref{SR}) close to minus infinity, then the asymptotic expanding space-time ${\mathcal E_{r}}$ is a space-time of singularities for all radius $0<r<r_0$.

ii) Based on (\S\ref{SE}., 1.) each basic element of the initial space ${\mathcal E_{i}}$ is a limit entity as the radius of each basic element of the primordial space-time ${\mathcal E_1}$ tends to zero. It gives to that limit entity a topology close to that of a point, and indicates that the basic elements of the initial space ${\mathcal E_{i}}$ are extremely contracted, with a local entropy $S_i$ given by the limit
\begin{equation}\label{Si}
S_i=\lim_{r\rightarrow0}-k\ \ln (\delta(r))=-\infty,
\end{equation}
\end{rmak}

\section{Interpretation of the earliest conditions and gravity origin}

 The above model presents the space-time of singularities ${\mathcal E_{r}}$, for all $0<r<r_0$ with $r_0\ll 1$, as the first state of decompression of an earliest extremely contracted limit space ${\mathcal E_{i}}$. It introduces the space-time of singularities as the first expanding space-time that quantifies the period of expansion between the extremely contracted space ${\mathcal E_{i}}$, and the primordial space-time ${\mathcal E_{1}}$ that approximatively last for 380 000 years after the Big Bang. The consideration of matter, energy and radiation within this model allows to deduce a possible scenario for the earliest phase of the space-time expansion. It involves extreme divergent conditions that incubate matter recombination and lead to trace back the gravity origin. Indeed, the model allows to establish a chronology of extreme transformation at the beginning of the space-time expansion as follow:

\subsection{Period of the initial space ${\mathcal E_{i}}$}

 The period of the initial space ${\mathcal E_{i}}$ is the period at which matter, energy and radiation are assumed to be assembled before the Big Bang (\S\ref{SE}, 1.) (it is the backward period in time of the primordial space-time ${\mathcal E_{1}}$). The initial space ${\mathcal E_i}$ represents an approximation of the space before the beginning of the expansion. It is a space endowed with a local entropy close to minus infinity (\ref{Si}) and a normal curvature close to plus infinity. It corresponds to an infinite space with extremely contracted basic elements endowed with a topology close to that of a point.

 \subsection{Period of the space of singularities ${\mathcal E_{r}}$ and gravity origin}

The period of the space of singularities is the closest period after the beginning of the space expansion (the Big Bang), quantified by a space-time of singularities  ${\mathcal E_{r}}$ for all $0<r<r_0\ll1$. This space-time of singularities is characterized by a normal curvature close to plus infinity, and an entropy close to minus infinity (Property \ref{ProSing}, Remark \ref{RMAK}). The consideration of matter and energy within this model leads to understand the major role of those earliest singularities for the recombination of matter and all form of energy after the Big Bang.\pesp

For this purpose we assume that the expansion of the initial space ${\mathcal E_{i}}$ starts at the minimal time $t_0$, when the dismantlement of matter at all scale begins via simultaneous expansion of the basic elements of ${\mathcal E_{i}}$. Therefore the minimal time $t_0$ marks the beginning of the period of the space-time of singularities that includes the following:

\subsubsection{Short period of inflation}\label{SPI}

Observing matter receding from each other in all direction is an indication that matter and all form of energy were assembled in the past within a volume of an extremely contracted space. This space is described within this model by the initial space ${\mathcal E_{i}}$. Thus any volume in the initial space ${\mathcal E_{i}}$ is described by an infinite accumulation of extremely contracted limit entities with a topology close to that of a point. If all extremely contracted limit entities expand simultaneously to a tiny spheres, the initial space ${\mathcal E_{i}}$ expands instantaneously during a short inflation. This short inflation is the first phase of the space expansion, induced by a simultaneous decompression of extremely compressed dense basic elements of the initial space ${\mathcal E_{i}}$. Any finite volume in the initial space ${\mathcal E_{i}}$ expands via simultaneous expansion  of its basic elements to an infinite space of stacked tiny spheres.\pesp

 Indeed the transformation of each basic element of the initial space ${\mathcal E_{i}}$ from an extremely contracted limit entity (with the topology close to that of a point) to a tiny sphere induces the following:

\begin{enumerate}
  \item An instantaneous dismantlement of matter and energy. The density of the initial space ${\mathcal E_{i}}$ induces matter dismantlement at all tiny scales as all extremely contracted limit entities expand simultaneously to tiny spheres pushing everything outside of its volume.

  \item An instantaneous short period of inflation. The short inflation induces an extremely rapid recession motion of all amount of the dismantled matter and energy from each other in all directions. It is a consequence of the simultaneous expansion of the space-time's basic elements from dimensionless basic elements to tiny basic elements of three dimensions. This process defines the mechanism that generates the instantaneous inflation.

  \item Any amount of the dismantled matter undertakes a local motion due of its instantaneous dismantlement at all scales (it becomes in a state of motion). This local motion is different from the recession motion of all amount of dismantled matter due to space expansion.

  \item The simultaneous expansion of the extremely contracted basic elements transforms the initial space ${\mathcal E_{i}}$ into the space-time of singularities $({\mathcal E_{r})}_{r>0}$ for all $0<r<r_0\ll1$. Any basic element of the transformed space-time $({\mathcal E_{r})}_{r>0}$ for all $0<r<r_0$ is characterized by a normal curvature close to plus infinity and an entropy close to minus infinity. The singularities of $({\mathcal E_{r})}_{r>0}$ for all $0<r<r_0\ll1$ are called primordial singularities.

  \item The motion of any amount of the dismantled matter within singularities induces a centripetal acceleration directed toward the center of each singularity. The induced centripetal acceleration is given by $\gamma_r=v^2 {\delta_r}$,
       where $v$ is the module of the local speed induced by matter dismantlement and ${\delta_r}$ is the singularity's normal curvature. This radial acceleration is responsible for the changes in the direction of the velocity of any amount of the dismantled matter to follow the normal curvature of the space. Since ${\delta_r}$ is close to plus infinity, then any amount of the dismantled matter experiences a radial acceleration close to plus infinity directed toward the center of the singularity where it was distributed. As the effect of gravity and acceleration are indistinguishable (principle of equivalence), then any amount of the dismantled matter experiences a local centripetal gravitational field close to plus infinity.

  \item Based on Newton's second law, a force must cause this centripetal acceleration of magnitude
  \begin{equation}\label{Gforce}
  F=mv^2\delta_r.
  \end{equation}
  This force bends the motion of any amount of the dismantled matter to the space normal curvature. It makes the curved motion possible in each singularity. Without this force, the motion of any amount of the dismantled matter would continue in straight line instead of following the normal curvature of the space. It is a force that imposes a bent motion adapted to the normal curvature of the space.

  \item The existence of a centripetal force close to plus infinity within each singularity traps any amount of the dismantled matter locally just after the beginning of its dismantlement. The simultaneous expansion of the basic elements of the initial space  ${\mathcal E_{i}}$ into singularities dismantles matter and energy at all scales, and traps locally any amount of the dismantled matter and energy within an infinite family of stacked tiny spheres. This process leads to an uniform distribution of the dismantled matter and energy in the space-time of singularities during a fraction of the first microsecond of expansion of the space ${\mathcal E_{i}}$.
\end{enumerate}

Moreover, the centripetal force close to plus infinity within each singularity traps any amount of the dismantled matter and energy and makes it experience an infinite local centripetal compression propitious to matter recombination. In particular it incubates the recombination of the elementary particles at the first phase and later on the simplest atomic nucleus.

\subsubsection{Gravity origin}

Before the beginning of the expansion,  matter and energy were assembled in an initial space ${\mathcal E_{i}}$. The simultaneous expansion of the dimensionless basic elements of ${\mathcal E_{i}}$ into basic elements of three dimensions that pushes everything outside their volume, transforms the initial space ${\mathcal E_{i}}$ into the space-time of singularities $({\mathcal E_{r})}_{0<r<r_0}$, for an arbitrary tiny $r_0$ (\S \ref{SPI}, item 4.). This first transformation induces the following:

\begin{itemize}
\item a non null momentum of any amount of the dismantled matter that defines its state of motion. It is a consequence of its transformation from assembled structures to a dismantled structures at all scales,
\item a local trap for any amount of the dismantled matter within singularities. It is a consequence of its state of motion within an extremely curved space-time,
\item an intensely accelerated recession motion of any trapped amount of the dismantled matter from each other in all directions. It is a consequence of the short inflation period during the first microsecond of the expansion of the initial space ${\mathcal E_{i}}$.
\end{itemize}

After the total dismantlement of matter and energy at all scales during the instantaneous inflation period, gravity finds its origin revealed according to the principle of equivalence between gravity and acceleration. Indeed, based on \S \ref{SPI} items 5. and 6., a centripetal acceleration close to plus infinity appears within any singularity of the space-time $({\mathcal E_{r})}_{0<r<r_0}$ where the dismantled matter and energy are distributed in a state of motion. According to the Newton's second law, a force must cause this centripetal acceleration. A centripetal force within each singularity of magnitude $F=mv^2\delta_r$, accelerates any amount of the dismantled matter and energy by changing its velocity direction without changing its speed, to make the bent motion adapted to the extremely curved singularity. Therefore any trapped amount of the dismantled matter and energy in a state of motion in the singularities experiences a radial gravitational pull inward the geometrical center of each singularity. This radial gravitational pull is responsible for the deviation of the motion of the trapped amount of the dismantled matter, from linear motion to curved motion, adhering to the extremely curved space-time.\pesp

The appearance of the centripetal gravitational force within each singularity is an indication of a state of motion of any trapped amount of the dismantled matter and energy in the singularity. It gives to the singularity the characteristics of a gravitational singularity that include a normal curvature close to plus infinity, a gravity close to plus infinity and an entropy close to minus infinity. Without this local centripetal force close to plus infinity, the local motion within each singularity of anything with mass would be impossible. The centripetal gravitational force is a kind of frictional force on any trapped amount of the dismantled matter from the expanding space-time to impose a bent motion adhering to its curvature. It reveals the resistance that any trapped amount of the dismantled matter offers to change the state of its motion direction from rectilinear to curvilinear motion. However, without the state of motion of the amount of the dismantled matter in the expanding space-time, the radial gravitational pull would not exist. This leads to assert the following: {\it gravity exists because matter is never at rest in a curved space-time}.

\subsubsection{Transformation of the space-time of singularities ${\mathcal E_{r}}$}

Based on this model, the period of the space-time of singularities $({\mathcal E_{r})}_{0<r<r_0}$, for an arbitrary tiny $r_0$, corresponds to the period estimated to 380 000 years after the Big Bang. During this period the basic elements of the space-time $({\mathcal E_{r})}_{0<r<r_0}$ are singularities of two kinds:

\begin{itemize}
  \item singularities without distribution of dismantled matter and energy. They are characterized by a normal curvature close to plus infinity and an entropy close to minus infinity
  \item singularities with distribution of an amount of the dismantled matter and energy in a state of motion. They are characterized by a normal curvature close to plus infinity, an entropy close to minus infinity and a gravity close to plus infinity. These singularities are called gravitational singularities.
\end{itemize}

The gravitational singularities are endowed with a centripetal gravitational force close to plus infinity that traps everything inside including light. As the space-time of singularities ${\mathcal E_{r}}$ continues to expand via simultaneous expansion of its basic elements by increasing each radius $r$ toward $r_0$, the normal curvature $\delta_r={1\over r}$ of the space-time of singularities $({\mathcal E_{r})}_{0<r<r_0}$ decreases gradually from plus infinity. Therefore the centripetal gravitational force (\ref{Gforce}) within each gravitational singularity reduces its extreme intensity gradually together with the space-time expansion.\pesp

Thus the process of gradual decrease of the centripetal gravitational force intensity induces a decompression of the gravitational singularities from plus infinity (\S \ref{SPI}), which incubates the recombination of the dismantled matter. In particular it recombines the elementary particles such as quarks, muons, leptons, electrons, and positrons, followed by the recombination of the composite particles such as protons, neutrons, other composite particles, and later on the simplest atomic nucleus. The decompression of the gravitational singularities continues at the rate of the space-time expansion until it reaches the intensity that allows nuclei to trap electrons and later on to form atoms. At this stage only simple atoms of neutral hydrogen are recombined, the remaining intensity of the centripetal gravitational force that allows electron to be trapped in the orbit of nucleus is insufficient for nuclear fusion.\pesp

 As the space-time expands, all singularities of the space-time $({\mathcal E_{r})}_{0<r<r_0}$ loose simultaneously their extreme characteristics gradually: the normal curvature of each singularity undergoes a gradual decrease from plus infinity, inducing a gradual decrease of the extreme intensity of the gravity from plus infinity and a gradual increase of the entropy from minus infinity. This process of continuous changes of the characteristics of the singularities lasts until the recombination of the simplest neutral hydrogen atom and the release of the light to travel for the first time as the radius $r$ of each singularity of ${\mathcal E_{r}}$ reaches the critical radius $r_0$. This makes the universe transparent and reveals the surface of the last scattering from where the cosmic microwave background comes. It indicates the end for the space-time of singularities $({\mathcal E_{r})}_{0<r<r_0}$ that precedes the beginning of the primordial space-time ${\mathcal E_1}$.\pesp

Moreover to conserve the density of the space-time of singularities $({\mathcal E_{r})}_{0<r<r_0}$, as it expands, new singularities of different sizes appear in the interstices of its stacked basic elements. The new singularities of different sizes form an Apollonian gasket (illustrated in Fig.\ref{Fig.3}) within each interstice to fill the gap. Each singularity looses its extreme characteristics as its radius expands to reach the critical radius $r_0$. All the singularities of the interstices experience a transformation similar to the transformation of the primordial singularities of $({\mathcal E_{r})}_{0<r<r_0}$.

\subsection{The period of primordial space-time ${\mathcal E_1}$}

The period of primordial space-time ${\mathcal E_1}$ is the period that comes after the end of the space-time of singularities $({\mathcal E_{r})}_{0<r<r_0}$, approximatively 380 000 years after the Big Bang. The primordial space-time ${\mathcal E_1}$ appears when the space-time of singularities ceases to exist as the radius of each basic element of ${\mathcal E_r}$ reaches the critical radius $r_0\ll1$.
The release of light to travel for the first time marks the beginning of this period, and makes the universe transparent for observation. When gravity reaches an intensity that allows the recombination of the simplest atoms of hydrogen and releases light, matter and energy appear to be spread all around the previous singularities' location within an uniform distribution to form the primordial observed space-time ${\mathcal E_1}$ (the CMB) at the Step 1.\pesp

The newly recombined matter within the space-time ${\mathcal E_1}$ experiences a centripetal gravitational force proportional to the space-time normal curvature of magnitude given by Newton's second law $F_1=mv^2\delta_1(r)$, to the extent that for all ${n\geq 1}$ the centripetal gravitational force is given by $F_n=mv^2\delta_n(r)$ within each basic element of the space-time ${\mathcal E_n}$ where matter is distributed. This gravitational force experiences a very slow decrease at a rate set by the expansion rate of the space-time. The centripetal gravitational force  within each basic element remains invariant for a short period of time since the normal curvature is imperceptible (Remark\ref{IMP}, i)), and decreases for a large period of time as the space-time ${\mathcal E_n}$ expands.\pesp

The interstices of the accumulated basic elements of the space-time ${\mathcal E_1}$ are the locations where new singularities of different sizes appear to fill the gaps. Each singularity lasts for a period approximatively greater than or equal to 380 000 years. The expansion of the different singularities of the interstices of the space-time ${\mathcal E_1}$ depends on the rate of expansion of its basic elements.
As the primordial space-time ${\mathcal E_1}$ expands, one can deduce briefly what will happen to the first recombined matter as the space-time expands (more details will be the subject of a separate paper). The major changes of matter transformation starting from the space-time ${\mathcal E_1}$ are generated by the singularities of the interstices at the rate of the space-time expansion. The appearance of the first stars, the recombination of heavier atoms and the formation of big structures are the product of the gravitational singularities of the interstices. Indeed, the singularities of the interstices of the primordial space-time ${\mathcal E_1}$ carry a potential of great interest as the space-time expands for two main characteristics:

\begin{itemize}
  \item The first main characteristic of the interstices resides in the continual appearance of other imbedded interstices between any three expanding stacked singularities of different sizes of the space-time ${\mathcal E_1}$. These repeating appearances generate an iteration of imbedded interstices for each $n\geq1$. Each interstice is a nursery of increasing stacked singularities of different sizes.

  \item The second main characteristic resides in the behavior of the interstices when matter is trapped inside their singularities. Indeed, each basic element of the space-time ${\mathcal E_1}$ touches twelve neighboring basic element, which creates eight interstices, filled with singularities of different sizes that surround each basic element of the space-time ${\mathcal E_1}$. Each interstice becomes filled with stacked gravitational singularities if anything with mass is trapped inside it. Therefore
      each interstice is endowed with a variety of extreme decompressions generated by the gradual decrease of the gravitational forces of extreme intensities (close to plus infinity). This makes each interstice a family of incubators that traps and transforms the first recombined matter. Moreover, the distribution of the interstices with gravitational singularities in the space-time induces a variation in the intensity of the gravitational field of the space-time. These variations are propitious to change with time any uniform distribution of matter and energy in the space-time ${\mathcal E_1}$ into lumpy and condense under the extreme gravitational pull of the gravitational singularities of the interstices. Indeed, with their different extreme intensities of compression, the gravitational singularities of the interstices present adequate conditions to trigger the incubation of grouped stars, the recombination of heavier particles, and thus they are the appropriate locations for the accumulation of clouds of gas, dust of matter, and grouped stars to form galaxies and the big structures we know today.
\end{itemize}
 If the freshly recombined neutral hydrogen atoms are trapped by the singularities of the interstices of the expanding space-time ${\mathcal E_1}$, the different extreme compressions inside those gravitational singularities are propitious for the recombination of heavier elements, as well as to overcome the Coulomb barrier. If they trap the neutral hydrogen, the gravitational singularities of ${\mathcal E_1}$ trigger the proton-proton chain needed for the nuclear fusion of neutral hydrogen atoms at the first stage to produce helium, energy, radiation, and give birth to the first big number of stars in the universe to end the period of expansion of the space-time ${\mathcal E_1}$ in darkness.

\section{Conclusion}

 A new definition of entropy is introduced using a model that simulates an expanding space-time compatible with the fundamental principle of cosmology. This entropy is obtained by mean of a state function that measures the variation of the space-time normal curvature consequence of its expansion. It leads to work out a new insight that unravels our understanding of the earliest conditions that last for a period estimated to 380 000 years after the Big Bang. This most mysterious period of the standard model of cosmology remains not accessible to direct observation and subject to hypothetical approaches or speculations.\pesp

The model used in this investigation provides a reduced schematic description of a phenomenon that account for the properties of the space-time expansion compatible with observation. It allows to investigate the space-time characteristics and to approach the earliest conditions that incubate matter and energy recombination after their total dismantlement. The earliest phase and conditions are found to be described by a space-time of singularities $(\mathcal{E}_r)$, for $r\in]0,r_0[$ endowed with basic elements of extreme characteristics in terms of normal curvature close to plus infinity, and entropy close to minus infinity. As the topology of each basic element of the space-time of singularities approaches the topology of that of a point, the model describes an extremely contracted initial space $\mathcal{E}_i$ where matter and energy are supposed to be assembled before the Big Bang event. \pesp

The simultaneous decompression of extremely compressed dense basic elements of the initial space $\mathcal{E}_i$ generates the space-time of singularities $(\mathcal{E}_r)$ for $r\in]0,r_0[$, after a short period of inflation. The presence of any amount of the dismantled matter in a state of motion within those singularities transforms them into gravitational singularities, characterized by a normal curvature close to plus infinity, a gravity close to plus infinity and an entropy close to minus infinity. This extreme environment traps everything with mass including light. It prevents any observation of the earliest extreme conditions that incubate matter recombination during a period quantified by the space-time $(\mathcal{E}_r)$ for $r\in]0,r_0[$. This period is estimated to approximatively last for 380 000 years after the Big Bang.\pesp

 The state of motion of the dismantled matter within an extremely curved singularity gives birth to an infinite gravitational attraction directed toward the center of the singularity, a center that represents an external location equidistant from  all positions in the singularity. This extreme gravitational force is a frictional force on the dismantled matter in a state of motion exerted by the normal curvature of the space-time of singularities to adjust its direction to the motion within the curved space. The centripetal acceleration exerted to anything with mass in a state of motion within each singularity of the space-time $\mathcal{E}_r$ divulges the origin of a gravity with an extreme intensity that compresses any trapped thing with mass. The gradual decrease of the intensity of the gravitational attraction induces a decompression at the rate of the space-time expansion that transforms each gravitational singularity into an incubator for matter recombination.\pesp

 In particular the gradual decompression maintains suitable conditions for the recombination of elementary particles at the first stage, followed by the recombination of the composite particles and later on the recombination of the simplest atomic nucleus. This process goes on at the rate of the space-time expansion, until the intensity of the decompression allows nuclei to trap electrons in their orbits and form the first simple neutral atoms of hydrogen. This process suggests that the amount of liberated energy we measure today when nucleus parts are disassembled by nuclear fission might be traced back, using the principle of conservation of energy, to the earliest gravitational extreme pressure that was born within the gravitational singularities of the young expanding space-time.\pesp

This case study allows to:

\begin{itemize}
\item study the change of a state function depending on the normal curvature of an expanding space-time compatible with the fundamental principle of cosmology, using an expansion quantified by a sequence of space-time ${\mathcal E}=({\mathcal E_n})_{n\geq1}$ that expands via  simultaneous expansion of its basic elements. The state function measures the local transformation of the space-time from a highly compressed space to a lower compressed space.

\item justify the existence of an instantaneous inflation that generates approximatively the uniform distribution of matter and all form of energy observed in the surface of the last scattering (CMB). This inflation is the first transformation of an initial extremely contracted space $\mathcal{E}_i$ to the expanding space-time of singularities $(\mathcal{E}_r)$ for $0<r<r_0$.

\item quantify, characterize and provide an interpretation of the hidden earliest phase closest to the beginning of the space expansion, that lasts approximately for 380 000 years.

\item trace back the earliest conditions propitious to the incubation of matter and energy recombination after their dismantlement by the Big Bang. It allows to understand the environment and process that lead to the birth of the first stars.

\item assert that gravity exists in a curved expanding space-time because matter is never at rest.
\end{itemize}

Within this model the energy that generates the simultaneous expansion of the stacked equal spheres is unknown. But it induces a simultaneous decompression of the extremely contracted dense basic elements of the initial space $\mathcal{E}_i$ that approximates the space before the Big Bang. The densest model of stacked equal spheres to cover a volume is found to represent approximatively 74$\%$ of the volume, which suggests that taking into account the density of the space-time, the family of stacked equal balls responsible for the expansion of the space-time within this model represents approximatively 74$\%$, and this estimation does not change with time.

\begin{small}

\end{small}

\end{document}